\documentclass[10pt,conference]{IEEEtran}
\IEEEoverridecommandlockouts
\usepackage{cite}
\usepackage{amsmath,amssymb,amsfonts}
\usepackage{algorithmic}
\usepackage{graphicx}
\usepackage{textcomp}
\usepackage{xcolor}
\usepackage{url}
\usepackage[T1]{fontenc}
\def\BibTeX{{\rm B\kern-.05em{\sc i\kern-.025em b}\kern-.08em
    T\kern-.1667em\lower.7ex\hbox{E}\kern-.125emX}}
    
\makeatletter
\newcommand{\linebreakand}{%
  \end{@IEEEauthorhalign}
  \hfill\mbox{}\par
  \mbox{}\hfill\begin{@IEEEauthorhalign}
}
\makeatother

\begin{document}

\title{Bluejay: A Cross-Tooling Audit Framework For Agile Software Teams
}

\author{
    \IEEEauthorblockN{C\'esar Garc\'ia}
    \IEEEauthorblockA{\textit{I3US Institute} \\
    \textit{Universidad de Sevilla}\\
    Sevilla, Spain \\
    cgpascual@us.es}
    \and
    \IEEEauthorblockN{Alejandro Guerrero}
    \IEEEauthorblockA{\textit{I3US Institute} \\
    \textit{Universidad de Sevilla}\\
    Sevilla, Spain \\
    aleguedia@alum.us.es}
    \and
    \IEEEauthorblockN{Joshua Zeitsoff}
    \IEEEauthorblockA{\textit{EECS Department} \\
    \textit{University of California}\\
    Berkeley, USA \\
    jzeitsoff@berkeley.edu}
    \and
    \IEEEauthorblockN{Srujay Korlakunta}
    \IEEEauthorblockA{\textit{EECS Department} \\
    \textit{University of California}\\
    Berkeley, USA \\
    srujay@berkeley.edu}
    \linebreakand
    \IEEEauthorblockN{Pablo Fernandez}
    \IEEEauthorblockA{\textit{SCORE Lab, I3US Institute} \\
    \textit{Universidad de Sevilla}\\
    Sevilla, Spain \\
    pablofm@us.es}
    \and
    \IEEEauthorblockN{Armando Fox}
    \IEEEauthorblockA{\textit{EECS Department} \\
    \textit{University of California}\\
    Berkeley, USA \\
    fox@berkeley.edu}
    \and
    \IEEEauthorblockN{Antonio Ruiz-Cort\'es}
    \IEEEauthorblockA{\textit{SCORE Lab, I3US Institute} \\
    \textit{Universidad de Sevilla}\\
    Sevilla, Spain \\
    aruiz@us.es}
}

\maketitle

\begin{abstract}
Agile software teams are expected to follow a number of specific
Team Practices (TPs) during each iteration, such as estimating the effort
(``points'') required to complete user stories and coordinating the
management of the codebase with the delivery of features.
For software engineering instructors trying to teach such TPs
to student teams,
manually auditing teams if teams are following the TPs
and improving over time is tedious, time-consuming and error-prone. It
is even more difficult when those TPs
involve two or more tools. For example, starting work on a feature in
a project-management tool such as Pivotal
Tracker should usually be followed relatively quickly by the creation
of a feature branch on GitHub. Merging a feature branch on GitHub
should usually be followed relatively quickly by deploying the new
feature to a staging server for customer feedback. Few systems are designed specifically to
audit such TPs, and existing ones, as far as we know, are limited to a
single specific tool.  

We present Bluejay, an open-source extensible platform that uses the APIs
of multiple tools to collect raw data, synthesize it into TP
measurements, and present dashboards to audit the TPs.
A key insight in Bluejay's design is that TPs can be expressed in
terminology similar to that used for modeling and auditing Service
Level Agreement (SLA) compliance. Bluejay therefore builds on mature
tools used in that ecosystem and adapts them for describing, auditing,
and reporting on TPs. 
Bluejay currently consumes data from five different widely-used
development tools, and can be customized by connecting it to any
service with a REST API. \\
Video showcase available at governify.io/showcase/bluejay
\end{abstract}


\section{Introduction}

\begin{figure}
    \centering
    \includegraphics[width=0.985\linewidth]{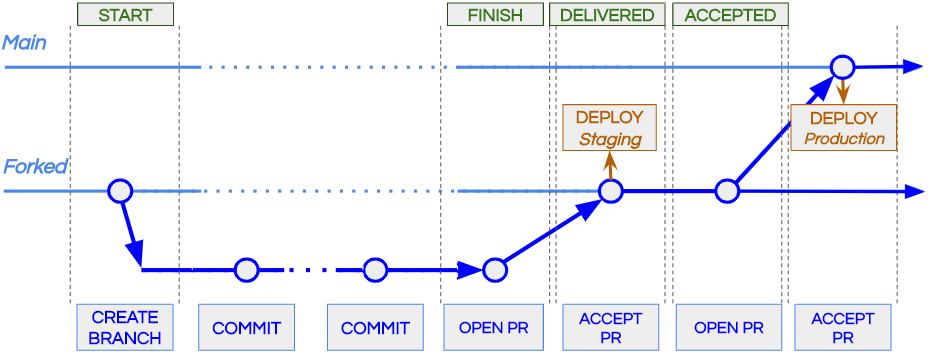}
    \caption{Multiple team workflow diagram. The boxes represent tool actions. Top green boxes correspond to changes in Pivotal Tracker story states, bottom blue boxes represent GitHub actions and the middle orange boxes correspond to deployment actions.}
    \label{fig:impr_analysis}
\end{figure}

Agile software development has become widespread in both industry and in the classroom.
Its lightweight processes and methodologies are increasingly well supported by external
tools; Figure
\ref{fig:impr_analysis} shows a typical workflow for projects
in which different teams contribute individually to a large code repository.
The upstream repository is forked and each planned feature is developed by
a member of the team in a separate branch. Once a feature is
completed, a pull request (PR) is opened for the whole team to discuss
the changes, and if accepted, merge the changes to the forked
repository's main branch. Once merged, the main branch is deployed to the
staging server for the client to validate new functionality. If the client signs off
on the feature, another PR is opened to merge the changes into the main
repository so the feature can be included in the production server. 

Many tools have emerged to support such workflows.
Project management tools such as  Pivotal Tracker (PT) allow users to
organize tasks and track who is working on which features.  Repository
management tools such as GitHub can be used to manage the code and
coordinate the merging of new code into the main line.
Continuous integration tools such as Travis CI run tests and linters 
to validate changes before accepting any
PR.  Deployment-as-a-service platforms such as Heroku even allow
automatic deployment from 
specific branches. 

Well established metrics for Agile processes rarely capture TPs that consist of sequences of actions \emph{across} such tools.
For example, when following a
 ``branch per feature'' development approach, a branch should be created shortly after a feature is claimed by a developer.
There exist multiple tools that integrate with GitHub or PT to help
audit teams such as ZenHub or Datadog but none of them are capable
of correlating multiple tool information. 
Therefore, both for Agile teams in industry and for instructors teaching these
methodologies, it is difficult to audit if teams are following practices that integrate multiple tools correctly.

Fortunately, there is already another domain
in which the challenge is to calculate performance indicators from
multiple services.  It is in the domain of Service Level Agreement (SLA)
management, and more specifically, expressing and evaluating Service
Level Objectives (SLOs). We find an important similarity between a
TP structure and SLO in SaaS.
For example, a typical SLO for an interactive Web site might be
\textit{``over any 5-minute window, 99\% of requests to the main page
  must be delivered within 200ms''}.  Analogously, a typical TP for
an Agile team might be \textit{``over any 2-week iteration, 75\% of
  stories should be '1-point' stories''}. In such a context, a set of
TP define a global agreement, that we coin as Team Practice
Agreement (TPA), in a similar way that a set of SLOs define a service
level agreement (SLA). Therefore we can leverage the models and tools that have already been successful for SLOs and adapt them for TPAs. 

Bluejay leverages the above insight for the challenge of multi-tool TPs.
It offers a systematic way to
define, measure and visualize TPs involving up to two
different services. The framework provides a tooling
ecosystem for instructors to define their own best practices to follow and
track the adherence of students teams in order to learn of
pitfalls and improve over time.  

Specifically, the framework provides a micro-services architecture
based on the Governify ecosystem~\cite{Muller2018} for SLA management. It has been extended in two different ways: a new Domain Specific Language (DSL) for adopting
different tools in order to create TPAs, and a customized dynamic
dashboard that shows the degree of adherence to practices over time
for the different teams. We have created components that connect to
popular existing Agile tools, but Bluejay is extensible both by adding
new external tools and by defining new metrics and practices based on
data from any supported external tools.

Section 2 introduces our system, including how to model, express, and audit Team
Practice Agreements (TPAs).
Section 3 describes our experience using the tool in real academic
scenarios.
Section 4 discusses related work.
Section 5 reviews our plans for future work.

\section{Bluejay}
 
\begin{figure}
    \centering
    \includegraphics[width=1\linewidth]{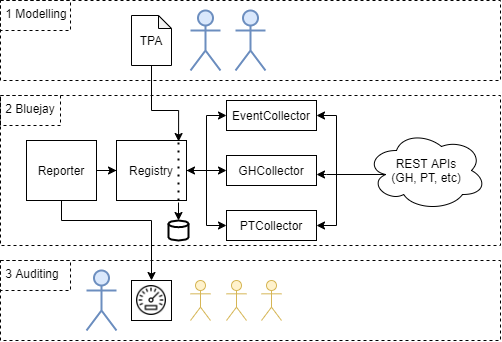}
    \caption{Bluejay schematic and usage diagram. Large blue actors represent the instructors and the smaller yellow actors are the students. }
    \label{fig:governify_diagram}
\end{figure}

\begin{figure*}
    \begin{tabular}{|p{3in}|p{3in}|c|}
    \hline
    Metric Pattern &
    Example TP &
    Objective     \\ \hline

    Number of [Event] in [Tool] every [Period] by [TEAM|MEMBER] &
    \textbf{TP\#1}: Number of OPEN\_PR in GITHUB every WEEK by MEMBER &
    $<= 1$  
    \\ \hline
    
    [MAX|MIN|AVG|STD|NEWEST|LATEST] [Property] value of [Event] in [Tool] every [Period] by TEAM  &
    \textbf{TP\#2}: AVG COVERAGE value of COVERAGE\_REPORT in  CODE by TEAM &
    $\leq 75\% $ 
    \\ \hline
    
    [Frequency] distribution of [Event] in [Tool] every [Period] by TEAM &
    \textbf{TP\#3}: DAILY distribution of SUCCESSFUL\_BUILDS in TRAVIS every WEEK by TEAM &
    $\sigma <= 1$ 
    \\ \hline

    Percentage of [Event1] in [Tool1] correlated with [Event2] in [Tool2] within [window] every [period] by TEAM &
    \textbf{TP\#4}: Percentage of NEW\_BRANCH (creation) in GITHUB correlated with START\_STORY in PIVOTAL\_TRACKER within 1\_HOUR every WEEK by TEAM  &
    $ \geq 75\% $ 
    \\ \hline
    \end{tabular}
    \caption{\label{doc:TPA_TABLE}
      Four metric patterns available for defining metrics
      of interest in Team Practices, and example Team Practices TP1--TP4 that
      follow each of the patterns.}
\end{figure*}

Bluejay is a framework composed of a set of components that allows modeling, monitoring, and auditing TPs.
In Figure \ref{fig:governify_diagram} a general vision of the process involving Bluejay is displayed.
The figure distinguishes a general workflow composed by three stages:

\begin{enumerate}
    \item First, \textbf{Modelling} is the process by which Instructors define a set of TPs, creating a TPA for teams to fulfill.
    \item Second, a monitoring process for each TPA is started; as part of this process \textbf{Bluejay} calculate the TPA fulfillment based on the metrics computed with the information gathered from the different REST APIs; the actual fulfillment information and metrics values are subsequently rendered in a graphical dashboard available for both instructors and teams.
    \item Third, Instructors can use the dashboard to \textbf{audit} teams and determine if they adhere to the practices.
\end{enumerate}

\subsection{TPA Modeling}
The user data input in Bluejay are the Team Practice Agreements, or TPAs. A TPA is composed of multiple TPs, each of which involves one or more metrics, and each metric is computed using the information from one or more tools. 

A Metric Pattern (MP) is a template for collecting, aggregating, measuring, and evaluating a raw metric from a specific tool.  An example of a raw metric might be a count from PT of the number of currently-active user stories in the team's project (stories that have been started but not completed). Another example might be the overall test coverage of the most recent CI run of the project, as reported by a tool such as Travis CI.
We identified four Metric Patterns that capture a variety of Team Practice behaviors useful in expressing TPAs. Figure \ref{doc:TPA_TABLE} shows the four metric patterns and an example of a Team Practice that uses each. We inspired in how process performance indicators are specified by using linguistic patterns\cite{LinguisticPatterns}. As the table shows, each TP is defined by a metric, an objective defining the valid value range for the metric, and a scope that determines whether the metric should be applied to the whole team or to each member separately. Each instructor determines which TPs they wish to audit for their students.  The tool's complete documentation details which raw metrics are available, how to define new TPs, and how to add new additional external tools' metrics and TPs.

\begin{figure}[b]
    \centering
    \includegraphics[width=1\linewidth]{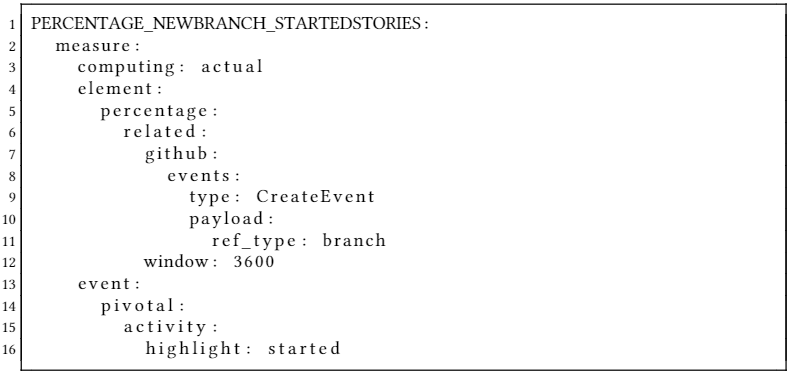}
    \caption{iAgree YAML syntax corresponding to TP\#4 in Figure~\ref{doc:TPA_TABLE}.}
    \label{doc:TP_PERCENTAGE}
\end{figure}

The TPA must be expressed in the syntax of iAgree, a YAML-based domain-specific language used in Governify.  Figure \ref{doc:TP_PERCENTAGE} shows the iAgree YAML corresponding to TP \#4 in the table.  The expression evaluates to  the percentage of started stories in PT that are correlated in time  with a branch creation event on GitHub for a given period of time. A catalogue of examples %
\cite{BluejayMetrics}
is available for public usage; later in the paper we describe how Bluejay can be extended with additional external tools, additional metrics, and additional metric patterns.

\subsection{TPA Monitoring}
Once TPAs are defined in iAgree, Bluejay can begin monitoring them.
The middle dotted box in Figure \ref{fig:governify_diagram} shows the main components of Bluejay.
The \textbf{Registry} stores TPAs in the iAgree language.
The \textbf{Reporter} is the component responsible for the main workflow; it coordinates the work of \textbf{collectors}, each of which is customized to retrieve data and compute metrics for a unique external tool.  A TPA and its TPs can use multiple collectors.
When the TPA is fully calculated, the reporter dynamically adapts the dashboard for auditing the team.
The initial implementation of Bluejay  has three collectors, but more can be added:

\begin{itemize}
\item \textbf{PTCollector}: Collects data from PT, for defining metrics related with PT stories for the team or for each individual member. E.g. ``One story started at a time for each member''
\item \textbf{GHCollector}:  Collects data from GitHub for defining metrics related to GitHub Projects boards. E.g. ''One card in "In progress" column at a time''.
\item \textbf{EventCollector}: A generic way to fetch data from 2 different APIs and correlate two different payloads. This is used to create complex TPs that more accurately represent the needs of auditing Agile teams using multi-tool workflows. Currently, the EventCollector is able to connect with 5 different data sources: PT, GitHub, Heroku, TravisCI and Codeclimate.
\end{itemize}

\subsection{TPA Auditing}
\begin{figure*}
    \centering
    \includegraphics[width=0.9\linewidth]{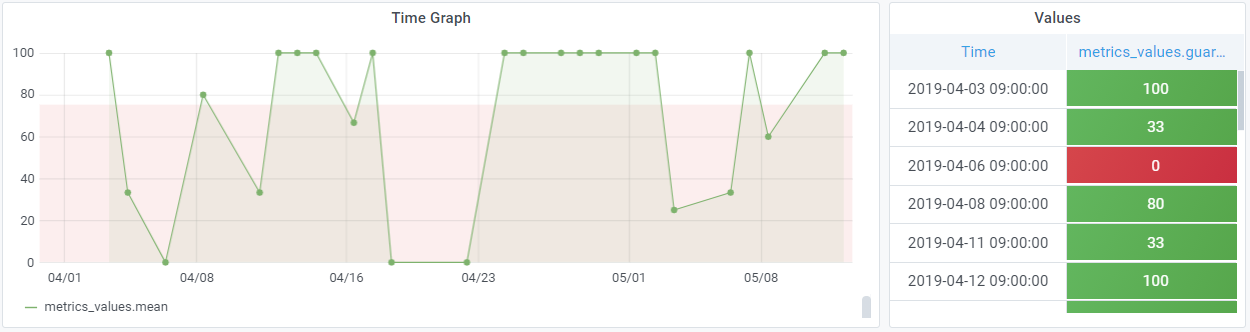}

    \includegraphics[width=0.9\linewidth]{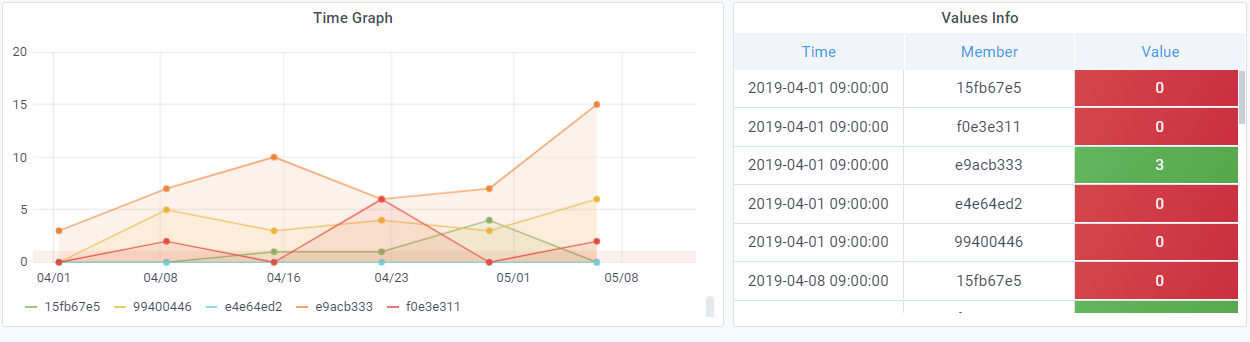}

    \caption{Top: Team auditing graph, showing the team's performance on TP\#4.  Bottom: Individual team member auditing graph, showing different team members' performance on TP\#1, which specifies that each individual member should at least open a PR each week to commit their work.}
    \label{fig:percentage_graph}
\end{figure*}

Bluejay  offers
diagrams to audit both
team-level TPAs and 
member-level TPAs.  The system will automatically generate a dashboard for each TPA based on the scope of each guarantee. 
Both 
graphs will appear as a pair of graph and table, containing the obtained values after computing all the guarantees. The X axis of the graph corresponds to the time and the Y axis to the actual value of the metric. The table contains the same points but displayed as a pair of date and value.
Figure \ref{fig:percentage_graph} shows a screenshot of a fragment of the dashboard including a graph representing TP\#4 from figure \ref{doc:TPA_TABLE}.

The top (team) graph in the figure shows that most of the time, the team creates a new feature branch in GitHub within 1 hour of starting a story on PT. On the right side, the actual value of the points for each date can be seen inside a green box if that point fulfills the objective or a red box if not.
The bottom (team members) graph shows
a colored line for each team member, and the table at the top shows each member's compliance with TP\#1, which may be stated as ``Each member should open at least one PR every week.''
Users can change the period of time to analyze. This is useful when looking for practice violations within a certain iteration.

\subsection{Customization and Extension}
\begin{figure} 
    \includegraphics[width=1\linewidth]{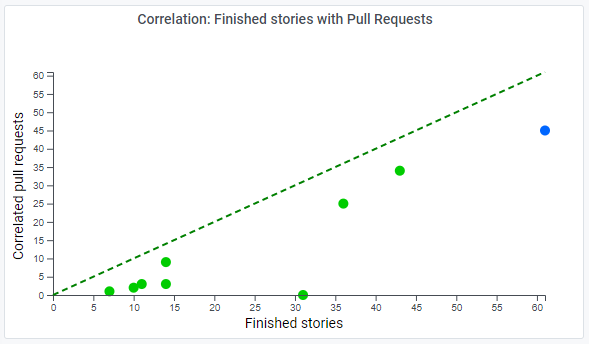}
    \caption{Finished stories in PT correlated with a opened PR on GitHub. Each point represents a team.}
    \label{fig:correlation_graph}
\end{figure}

The patterns described in the second section categorize the metrics that can be used currently in the system. The system is prepared to be extended in five ways:

\begin{itemize}
    \item \textbf{TPAs} can be created and extended to audit all the TPs certain team needs. There is an available catalogue%
    \cite{BluejayTPAs}
    of examples to use as starting point.
    \item \textbf{Metric}: As the Event Collector can combine information from multiple tools or extract information in a generic way, new metrics can be created by modifying the DSL. We also have an available metrics catalogue%
    \cite{BluejayMetrics}.
    \item \textbf{Data sources}: Any new data source featuring a REST API, such as Trello, can be included for usage in the Event collector. There exists a detailed walk-through%
    \cite{BluejayCollectorExtension}
    \item \textbf{Collectors}: The existing collectors can be modified to include new metric types and define new patterns.
    In order to create and integrate more collectors (e.g. SOAP collector) a walkthrough and template is available
    \cite{BluejayCollectorTemplate}.
    
    
    \item \textbf{Dashboard}: Bluejay also allows users to create specific graphs to customize the dashboard. In Figure \ref{fig:correlation_graph} a graph showing the level of correlation between PT stories and GitHub PRs is displayed. PT allows users to link GH branches and PR inside the stories to link them with their corresponding feature branch. In the X axis the number of finished stories per team are represented. The Y axis shows the number of PR which are linked with a story in PT. The closer the team is to the y=x line, a higher percentage of stories are correlated with PRs. Each point represents a team. More information about customized dashboards can be found on Bluejay's documentation 
    \cite{BluejayCustomGraphs}.
\end{itemize}

As an adaptation example, if a instructor needs to integrate Trello for auditing teams, if the metric follows any pattern of the presented in Figure \ref{doc:TPA_TABLE}, the only requirement is to extend the event collector to include the new data sources for the Trello API. If the metric is not compatible with the given patterns there are two options: extending the event collector to support the new metric, or creating a new collector particularly for the specific type of metric.

\section{Impact discussion}
The system has been applied in two Software Engineering teaching scenarios, in Spring 2019 and Summer 2020.

\textbf{Spring 2019.} In a software engineering course, 
      120 students were divided into 20 groups of six, working with a real customer
      on a web service project with 4 iterations of 2 weeks each. In this
      case study, we focus on backlog delivery practices. We would like to understand:
    \begin{itemize}
        \item Can TPAs change student backlog delivery behaviors?
        \item Can Bluejay change student backlog delivery behaviors?
    \end{itemize}
    The TPAs were given to students in natural language in the second
    iteration and the tool was made available to students in the fourth iteration.

    \textbf{\textit{Findings.}}
    Students reported better backlog delivery behaviors in iteration 3
    and 4. The tool reported better adherence to the practice during
    iteration 2 when the TPs where given to the students. During the
    3rd iteration there was a small reduction in adherence, which could
    be caused by a fatigue phenomenon. Finally in the iteration 4 the
    tool reported the highest adherence, when the dashboard was
    available to students. 


 \textbf{Summer 2020.} We analyzed projects from a Agile
      software engineering summer course in 2020. Groups of one, two
      or three students formed a total of 31 teams. Teams had a single
      ten-day sprint to complete all the features. 
    We examined if students followed each of the following three pairs of events
    \begin{itemize}
        \item How often does starting a story correspond with creating a branch?
        \item How often does finishing a story correspond with opening a PR?
        \item How often does delivering a story correspond with merging a PR?
    \end{itemize}
    Each team had a Github repository and a Pivotal Tracker project to track progress on the features, and both services were audited with  Bluejay. They also were taught to integrate GitHub and PT in order to correlate those events.

    \textbf{\textit{Findings.}}
    In the ideal version of our workflow, each story is associated
    with exactly one branch, the branch is created when work on the
    story is started and the branch is merged when the story is
    delivered. In the practice we found that this correspondence is
    not always one-to-one and some teams did not
    connect their “started” stories to its branch; also,  many stories that were marked as ``finished'' but not  associated with an opened PR possibly due to a number of reasons: they never connect the story with a branch so there wouldn't be one to find a PR
    from, or, a branch might have been merged without opening a PR to solicit review.  Finally, many stories were marked as ``delivered'' but were not connected with a merged (closed) PR. This could be for at least two reasons.  First, ``delivered''
    stories may have had PRs opened for them that could not be
    merged before the course ended (students ran out of time), and second, for reasons similar to ``finished'' stories not being connected with open PR. 

\section{Related work}


To the best of our knowledge, currently no related tooling has been proposed in the literature. In \cite{10.1145/3338906.3341181}
, we proposed an initial prototype of the system for auditing team practices but it only
allow the usage of metrics in an isolated context lacking the cross-tooling correlation (e.g. TP\#4) that represents
 a fundamental feature to analyse the flow of the overall process. 

However, related work around professional software teams
defines software development practices that could be
defined as TPs in order to be measured and analyzed with our proposed
tool. Thus, Baltes et al.\ propose in \cite{Baltes_2018} a theory
describing software developers' properties and practices that are
distinctive for experts in their field. For example, the concept of
getting feedback from peers is identified as an important factor that
could be used as a metric in our TPs. Huijgens et
al.~\cite{Huijgens_2017} provide metrics with high predictive power towards a
subset of lagging variables. Thus, as part of their results, metrics
such as \textit{``planned stories completion ratio''} or \textit{``planned
  points completion ratio''} could be defined as TPs in our work,
as the authors ascribe to them a high predictive power. Treude et
al.~\cite{Treude_2015} and Meyer et al.~\cite{Meyer_2014}
conducted empirical studies with 156 GitHub users and 379 software
developers.  From their studies metrics emerged for
both (1) analysis for objective measures of development activity
and (2)  improvement of productivity through the development of
new tools and the sharing of best practices.  Bluejay
could be extended in this direction to incorporate such a set of new
metrics. Finally, Kupiainen et al.~\cite{Kupiainen_2014}
establish a motivation for our work by providing a
systematic literature review analyzing why and how metrics are used by
industrial agile teams in order to infer enforcing process
improvements. 

In software engineering education, there are recent studies on
exploiting metrics and dashboards to provide fast
feedback.  Matthies et
al.~\cite{matthies2016surveys} show how to use a metric
dashboard, ScrumLint, to provide fast feedback to students.  Bai et
al. use metrics to deliver continuous feedback to students in a
software engineering course~\cite{bai2018continuous}.
Our approach generalizes these techniques and provides an extensible
framework that integrates with external tools for capturing these and
other Team Practice metrics that instructors design. 

One challenge instructors of project-based Agile courses face is ensuring that students gain an understanding of core Agile concepts. Steghöfer et al.~\cite{Steghofer2016challenge} found that students felt delivering a final product was more important than understanding Agile processes. Bluejay can help instructors measure student adherence to Agile processes by describing the Agile practices taught in class as TPAs. Based on the scores for each TPA, an instructor can quickly come to a decision about which Agile practices are not being followed and intervene with the appropriate student or team.

\section{Conclusions and Future Work}

Bluejay leverages the existing Service Level Agreement (SLA) management
platform Governify\cite{Muller2018}, extended with a DSL for
Agile software development metrics and an interactive dynamic
dashboard. The result is automated auditing of team practices widely used in agile 
methodologies, especially when those practices involve 
computing metrics from data aggregated from multiple tools supporting Agile 
workflows. As future work we plan to extend the infrastructure adding
more tools and metric patterns, and to continue validating the system in
software teams where we can mine interesting TPs from historic data
and also audit new teams. 

Beyond an Agile context, Bluejay opens the possibility to a more systematic
approach for supporting software engineering teaching, first with a
way of defining a set of team practices for the teams to follow and
also when observing the results archived for a continuous
improvement. It is not easy to do this process manually and our tool
is able to automate it.  Bluejay can also be used
for software engineering teaching experimentation, for
example, defining a set of potential best practices and then observing
how teams' results are affected if they follow these practices. 

More information and installation instructions for Bluejay may be
found at: \url{https://governify.io/quickstart/auditing-agile}

\bibliographystyle{IEEEtran}
\bibliography{bibliography}

\begin{thebibliography}{10}
\providecommand{\url}[1]{#1}
\csname url@samestyle\endcsname
\providecommand{\newblock}{\relax}
\providecommand{\bibinfo}[2]{#2}
\providecommand{\BIBentrySTDinterwordspacing}{\spaceskip=0pt\relax}
\providecommand{\BIBentryALTinterwordstretchfactor}{4}
\providecommand{\BIBentryALTinterwordspacing}{\spaceskip=\fontdimen2\font plus
\BIBentryALTinterwordstretchfactor\fontdimen3\font minus
  \fontdimen4\font\relax}
\providecommand{\BIBforeignlanguage}[2]{{%
\expandafter\ifx\csname l@#1\endcsname\relax
\typeout{** WARNING: IEEEtran.bst: No hyphenation pattern has been}%
\typeout{** loaded for the language `#1'. Using the pattern for}%
\typeout{** the default language instead.}%
\else
\language=\csname l@#1\endcsname
\fi
#2}}
\providecommand{\BIBdecl}{\relax}
\BIBdecl

\bibitem{Muller2018}
C.~Müller, P.~Fernandez, A.~M. Gutierrez, O.~Martín-Díaz, M.~Resinas, and
  A.~Ruiz-Cortés, ``Automated validation of compensable slas,'' \emph{IEEE
  Transactions on Services Computing}, 2018.

\bibitem{LinguisticPatterns}
A.~del Río-Ortega, M.~Resinas, A.~Durán, and A.~Ruiz-Cortés, ``Using
  templates and linguistic patterns to define process performance indicators,''
  \emph{Enterprise Information Systems}, vol. 10:2, pp. 159--192, 2016.

\bibitem{BluejayMetrics}
\BIBentryALTinterwordspacing
G.~César, G.~Alejandro, Z.~Joshua, S.~Korlakunta, F.~Pablo, F.~Armando, and
  A.~Ruiz-Cortés. Bluejay: Metrics' catalogue. [Online]. Available:
  \url{https://github.com/isa-group/governify-examples/tree/master/metrics/event-collector}
\BIBentrySTDinterwordspacing

\bibitem{BluejayTPAs}
\BIBentryALTinterwordspacing
------. Bluejay: Tpas' catalogue. [Online]. Available:
  \url{https://github.com/isa-group/governify-examples/tree/master/TPAs/bluejay}
\BIBentrySTDinterwordspacing

\bibitem{BluejayCollectorExtension}
\BIBentryALTinterwordspacing
------. Bluejay: How to extend event collector's sources. [Online]. Available:
  \url{https://github.com/isa-group/governify-collector-events/wiki/How-to-increment-collector-with-new-API}
\BIBentrySTDinterwordspacing

\bibitem{BluejayCollectorTemplate}
\BIBentryALTinterwordspacing
------. Bluejay: Collector template. [Online]. Available:
  \url{https://github.com/isa-group/governify-collector-v2-template}
\BIBentrySTDinterwordspacing

\bibitem{BluejayCustomGraphs}
\BIBentryALTinterwordspacing
------. Bluejay: Custom graphs. [Online]. Available:
  \url{https://governify.io/customization/dashboards}
\BIBentrySTDinterwordspacing

\bibitem{10.1145/3338906.3341181}
\BIBentryALTinterwordspacing
A.~Guerrero, R.~Fresno, A.~Ju, A.~Fox, P.~Fernandez, C.~Muller, and
  A.~Ruiz-Cort\'{e}s, ``Eagle: A team practices audit framework for agile
  software development,'' in \emph{Proceedings of the 2019 27th ACM Joint
  Meeting on European Software Engineering Conference and Symposium on the
  Foundations of Software Engineering}, ser. ESEC/FSE 2019.\hskip 1em plus
  0.5em minus 0.4em\relax New York, NY, USA: ACM, 2019, p. 1139–1143.
  [Online]. Available: \url{https://doi.org/10.1145/3338906.3341181}
\BIBentrySTDinterwordspacing

\bibitem{Baltes_2018}
\BIBentryALTinterwordspacing
S.~Baltes and S.~Diehl, ``Towards a theory of software development expertise,''
  in \emph{Proceedings of the 2018 26th ACM Joint Meeting on European Software
  Engineering Conference and Symposium on the Foundations of Software
  Engineering}, ser. ESEC/FSE 2018.\hskip 1em plus 0.5em minus 0.4em\relax New
  York, NY, USA: ACM, 2018, pp. 187--200. [Online]. Available:
  \url{http://doi.acm.org/10.1145/3236024.3236061}
\BIBentrySTDinterwordspacing

\bibitem{Huijgens_2017}
\BIBentryALTinterwordspacing
H.~Huijgens, R.~Lamping, D.~Stevens, H.~Rothengatter, G.~Gousios, and
  D.~Romano, ``Strong agile metrics: Mining log data to determine predictive
  power of software metrics for continuous delivery teams,'' in
  \emph{Proceedings of the 2017 11th Joint Meeting on Foundations of Software
  Engineering}, ser. ESEC/FSE 2017.\hskip 1em plus 0.5em minus 0.4em\relax New
  York, NY, USA: ACM, 2017, pp. 866--871. [Online]. Available:
  \url{http://doi.acm.org/10.1145/3106237.3117779}
\BIBentrySTDinterwordspacing

\bibitem{Treude_2015}
\BIBentryALTinterwordspacing
C.~Treude, F.~Figueira~Filho, and U.~Kulesza, ``Summarizing and measuring
  development activity,'' in \emph{Proceedings of the 2015 10th Joint Meeting
  on Foundations of Software Engineering}, ser. ESEC/FSE 2015.\hskip 1em plus
  0.5em minus 0.4em\relax New York, NY, USA: ACM, 2015, pp. 625--636. [Online].
  Available: \url{http://doi.acm.org/10.1145/2786805.2786827}
\BIBentrySTDinterwordspacing

\bibitem{Meyer_2014}
\BIBentryALTinterwordspacing
A.~N. Meyer, T.~Fritz, G.~C. Murphy, and T.~Zimmermann, ``Software developers'
  perceptions of productivity,'' in \emph{Proceedings of the 22Nd ACM SIGSOFT
  International Symposium on Foundations of Software Engineering}, ser. FSE
  2014.\hskip 1em plus 0.5em minus 0.4em\relax New York, NY, USA: ACM, 2014,
  pp. 19--29. [Online]. Available:
  \url{http://doi.acm.org/10.1145/2635868.2635892}
\BIBentrySTDinterwordspacing

\bibitem{Kupiainen_2014}
\BIBentryALTinterwordspacing
E.~Kupiainen, M.~V. M\"{a}ntyl\"{a}, and J.~Itkonen, ``Why are industrial agile
  teams using metrics and how do they use them?'' in \emph{Proceedings of the
  5th International Workshop on Emerging Trends in Software Metrics}, ser.
  WETSoM 2014.\hskip 1em plus 0.5em minus 0.4em\relax New York, NY, USA: ACM,
  2014, pp. 23--29. [Online]. Available:
  \url{http://doi.acm.org/10.1145/2593868.2593873}
\BIBentrySTDinterwordspacing

\bibitem{matthies2016surveys}
C.~Matthies, T.~Kowark, K.~Richly, M.~Uflacker, and H.~Plattner, ``How surveys,
  tutors and software help to assess scrum adoption in a classroom software
  engineering project,'' in \emph{2016 IEEE/ACM 38th International Conference
  on Software Engineering Companion (ICSE-C)}.\hskip 1em plus 0.5em minus
  0.4em\relax IEEE, 2016, pp. 313--322.

\bibitem{bai2018continuous}
X.~Bai, M.~Li, D.~Pei, S.~Li, and D.~Ye, ``Continuous delivery of personalized
  assessment and feedback in agile software engineering projects,'' in
  \emph{2018 IEEE/ACM 40th International Conference on Software Engineering:
  Software Engineering Education and Training (ICSE-SEET)}.\hskip 1em plus
  0.5em minus 0.4em\relax IEEE, 2018, pp. 58--67.

\bibitem{Steghofer2016challenge}
J.~{Steghöfer}, E.~{Knauss}, E.~{Alégroth}, I.~{Hammouda}, H.~{Burden}, and
  M.~{Ericsson}, ``Teaching agile - addressing the conflict between project
  delivery and application of agile methods,'' in \emph{2016 IEEE/ACM 38th
  International Conference on Software Engineering Companion (ICSE-C)}, 2016,
  pp. 303--312.

\end{thebibliography}

\end{document}